\documentclass[10pt,aps,prl,twocolumn,superscriptaddress]{revtex4}
\usepackage[linktocpage=true, colorlinks=true, urlcolor=blue, linkcolor=blue, citecolor=blue]{hyperref}
\usepackage{graphicx}
\usepackage{amsmath,amssymb}
\usepackage{color}
\usepackage{subfigure}
\usepackage{url}

\graphicspath{{../Images/}}

\begin{document}

\title{Swept-source optical coherence tomography by off-axis Fresnel transform digital holography with an output throughput of 10 Giga voxels per second in real-time}

\author{E. Charpentier}
\author{F. Lapeyre}
\author{J. Gautier}
\author{L. Waszczuk}
\author{J. Rivet}
\affiliation{
Centre National de la Recherche Scientifique (CNRS) UMR 7587, Institut Langevin. Paris Sciences et Lettres (PSL) University, Universit\'e Pierre et Marie Curie (UPMC), Universit\'e Paris 7. \'Ecole Sup\'erieure de Physique et de Chimie Industrielles ESPCI Paris - 1 rue Jussieu. 75005 Paris. France
}
\author{S. Meimon}
\affiliation{
ONERA – the French Aerospace Lab, Ch\^atillon, France
}
\author{L. Puyo}
\author{J.P. Huignard}
\affiliation{
Centre National de la Recherche Scientifique (CNRS) UMR 7587, Institut Langevin. Paris Sciences et Lettres (PSL) University, Universit\'e Pierre et Marie Curie (UPMC), Universit\'e Paris 7. \'Ecole Sup\'erieure de Physique et de Chimie Industrielles ESPCI Paris - 1 rue Jussieu. 75005 Paris. France
}
\author{M. Atlan}
\affiliation{
Centre National de la Recherche Scientifique (CNRS) UMR 7587, Institut Langevin. Paris Sciences et Lettres (PSL) University, Universit\'e Pierre et Marie Curie (UPMC), Universit\'e Paris 7. \'Ecole Sup\'erieure de Physique et de Chimie Industrielles ESPCI Paris - 1 rue Jussieu. 75005 Paris. France
}

\date{\today}
%\date{}

\begin{abstract}
We demonstrate swept-source optical coherence tomography in real-time by high throughput digital Fresnel hologram rendering from optically-acquired interferograms with a high-speed camera. The interferogram stream is spatially rescaled with respect to wavelength to compensate for field-of-view dilation inherent to discrete Fresnel transformation. Holograms are calculated from an input stream of 16-bit, 1024-by-1024-pixel interferograms recorded at up to 512 frames per second with a digital camera. All calculations are performed by a NVIDIA TITAN Xp graphics card on single-precision floating-point complex-valued arrays (32-bit per quadrature). It allows sustained computation of 1024-by-1024-by-256-voxel volumes at 10 billion voxel/s, from which three perpendicular cuts are displayed in real-time at user-selected locations, up to 38 frames per second.
\end{abstract}

\maketitle
%%%%%%%%%%%%

Optical coherence tomography (OCT) refers to a wide range of non-invasive high resolution optical interferometric imaging technologies which have found major applications in ophthalmology. Point-scanning (flying spot) OCT allows for confocal gating of the detected photons, filtering of the out-of-focus light. However, nowadays, shot noise-limited point scanning confocal OCT modalities~\cite{FechtigKumar2014} have reached a throughput limit due to the maximum permissible exposure in ophthalmic applications. The high complexity of OCT systems that implement adaptive optics \cite{KocaogluTurner2014} hinders their wide spread use in clinical research. Full-field swept-source OCT (FF-SS-OCT) is an OCT implementation that can acquire data several orders of magnitude faster with a sensor array. FF-SS-OCT allows for acquisition of volumes in a single laser sweep, eliminating artifacts from imperfect phase reproducibility during laser sweeps~\cite{Hillmann2016}. Finally, FF-SS-OCT allows for off-axis schemes, that improve image quality by spatial filtering of interferometric terms~\cite{Hillmann2012}. Line-field OCT is a hybrid method, benefiting from the advantages of confocal and full field detection schemes, while suffering from their respective limitations. Images are recorded with a single line camera; a lateral scanning and the laser sweep allow for three-dimensionnal imaging. The partial confocal filter of line-field OCT permits high quality and high-speed imaging at increased illumination levels with respect to bidimensional scanning schemes~\cite{FechtigGrajciar2015}. With a setup complexity slightly lower than its confocal counterparts, line-scanning OCT offers good promises of high image quality through digital refocusing of in-depth acquisitions~\cite{FechtigKumar2014, GinnerSchmoll2018}.

Imaging speed is crucial in optical coherence tomography (OCT) systems for imaging of dynamic samples~\cite{KleinHuber2017}. The presence of motion artefacts deteriorates the resolution of the OCT images. Consequently, acquisition speed and spatial resolution are ultimately linked~\cite{PovazayHofer2009}. Holographic OCT with a swept-source laser on a sensor array can acquire data several orders of magnitude faster than scanning OCT modalities~\cite{Hillmann2016}. As the technology of high-speed cameras, wavelength-swept lasers, and parallel computing improve, holographic OCT should become a cost-effective alternative to complex OCT systems that implement adaptive optics~\cite{KocaogluTurner2014}. The main advantages of holographic OCT lie in: 1- a reduced system complexity, in which no moving parts are involved, 2- the available detection throughput of high-speed CMOS cameras, 3- the possibility to perform aberration correction in post-processing~\cite{Hillmann2016}, 4- the computation of 3D motion fields~\cite{SpahrPfaffle2018}. %For eye fundus imaging, the wavelength scan should be performed within less than 5 ms, to circumvent axial and lateral motion artifacts~\cite{Bonin2010}. This sets order of magnitude of the lower required imaging frame rate to $\sim$ 100 kHz, for the acquisition of $\sim$ 500 images needed for tomographic reconstruction in $\sim$ 5 ms, which state-of-the-art high-speed cameras are now able to sustain.\\
A fast, robust and versatile digital image acquisition and rendering software is a key requirement for the development of digital holographic imaging. Video-rate holographic image rendering was made possible by streamline processing of optically-acquired interferograms on graphics processing units (GPUs)~\cite{ShimobabaSato2008, Ahrenberg2009}. When real-time performance is required, hardware acceleration of computations is an efficient way of increasing the throughput \cite{GaoKemao2012, Reid2012}. Special-purpose field-programmable gate array (FPGA) chips were reportedly used for high throughput holographic image rendering~\cite{Kakue2015}, at the price of less versatility than with general-purpose graphics processing units. Here, we demonstrate digital hologram rendering for swept-source optical coherence tomography with an output throughput of 10 billion voxels per second.

\begin{figure}[]
\centering
\includegraphics[width = 8.0 cm]{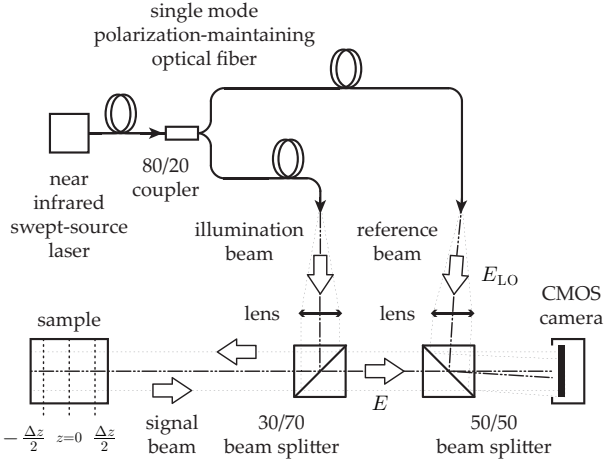}
\caption{Sketch of the wavelength-tuning laser holographic interferometer prototype used for demonstration.}
\label{fig_Setup}
\end{figure} 
%and wavenumber $k = \omega_{\rm L}/c$

Digitized interferograms were measured with the setup sketched in Fig.~\ref{fig_Setup}. It is an off-axis (angle $\sim 1 ^\circ$), interferometer used for optical detection of an object field $E$ beating against a separate local oscillator field $E_{\rm LO}$, in reflective geometry. A tunable laser (Broadsweeper BS-840-2-HP from Superlum) emits a 40 mW radiation whose angular optical frequency $\omega$ is swept linearly with time from $\omega_1$ to $\omega_2$, during a sweep time $T = 2\pi/\omega_{\rm sweep}$ = 1 s. These bounds are linked to the start wavelength $\lambda_1 = 870 \pm 2$ nm and the end wavelength $\lambda_2 = 820 \pm 2$ nm respectively via the relation $\omega = 2\pi c/\lambda$, where $c$ is the speed of light, and define a tuning range $\Delta\lambda = |\lambda_2 - \lambda_1|$ = 50 nm. As a consequence, the angular wavenumber $k = 2\pi/\lambda$ of the laser radiation is swept linearly with time. Interference patterns were recorded with a Adimec Quartz camera (pixel size $d = 12 \, \mu \rm m$), via a CoaXPress framegrabber Bitflow Cyton CXP4, at a frame rate of $\nu_{\rm S} = \omega_{\rm S} / (2 \pi) = 256 \, \rm Hz$, with 16-bit/pixel quantization. Each raw digitized interferogram of $1024 \times 1024$ pixels at position ($x,y$) and time $t$ is noted $I(x,y,t) =  \left| E + E_{\rm LO} \right| ^2$, where the object field is $E = {\cal E} \exp(i\omega t)$, the reference field is $E_{\rm LO} = {\cal E}_{\rm LO} \exp(i\omega t-i\phi)$, and $\phi$ is the phase detuning between both optical fields. Off-axis configuration allows for spatial separation of the self-beating $|E|^2$, and $|E_{\rm LO}|^2$ and cross-beating interferometric contribution, $E E_{\rm LO}^* = {\cal E} {\cal E}_{\rm LO}^* \exp(i\phi)$, and its complex conjugate~\cite{LeithUpatnieks1962}.

%In the linear response approximation, let $h(t)$ be the complex reflectivity of the sample at depth $z=ct$. The cross term of the interferogram takes the form of the cross-correlation between the backscattered field $E'=E*h$ and the original field $E$~\cite{Fuji1997, FercherDrexler2003}. The beating frequency spectrum of this term is the product between the spectrum of the radiation and the transfer function $WSR(\omega) = \tilde\Gamma(\omega) \tilde h(\omega)$.

%The use of \textbf{coherent} wavelength-tuning in radar, sonar, followed by laser interferometry ~\cite{Eickhoff1981, fercher1995} for ranging applications served as foundation for the development of swept-source OCT schemes~\cite{PanBirngruber1995}. 

\begin{figure}[]
\centering
\includegraphics[width = 8.0 cm]{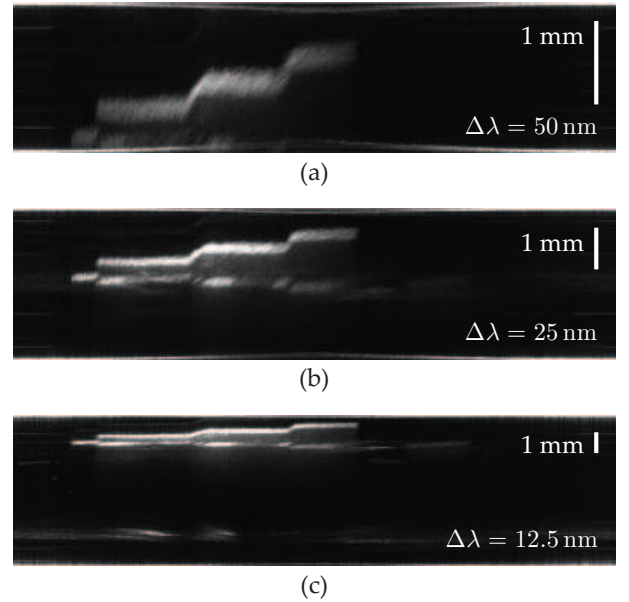}
\caption{Axial range $\Delta z$ vs. wavelength tuning range $\Delta\lambda$ of 50 nm (a), 25 nm (b), and 12.5 nm (c). The height difference between adjacent scales is $\sim 313 \, \mu \rm m$, which was measured by full-field OCT. The axial field of view $\Delta z$ of 1.8 mm (a), 3.7 mm (b), and 7.3 mm (c) is in agreement with the theoretical value from Eq.~\ref{eq_AxialRange} for $N_z = 256$. Axial scalebar : 1 mm.}
\label{fig_ScalesOCT}
\end{figure}

In swept-source OCT, a linear variation of the instantaneous optical angular frequency $\omega_{\rm L} = \beta t$ with time $t$ during a sweep results in a phase variation $\phi = 2kz$ of the interferometric beat between the wave backscattered by a diffuser at axial position $z$ and the reference wave
\begin{equation}\label{eq_PhaseDelay}
\phi = \frac{2\beta z}{c}t
\end{equation}
of the cross-beating part of the interferogram scaling up linearly with the sweep speed $\beta = (\omega_2 - \omega_1)/T$ of the laser angular frequency and with the detuning pathlength $z$ between the object and reference waves~\cite{FercherDrexler2003}. Hence, the instantaneous beat frequency 
\begin{equation}\label{eq_BeatingFrequencyVsPathlength}
\omega = \frac{\partial \phi}{\partial t} = \frac{2\beta z}{c}
\end{equation}
of the interferogram encodes the optical reflectivity signal at a detuning pathlength $z$ of the interferometer. Hence the Fourier transform of the temporal trace of the beat signal scales as the optical reflectivity against $z$~\cite{FercherDrexler2003}. The total restituted axial range $\Delta z$ can be derived from Eq.~\ref{eq_BeatingFrequencyVsPathlength} and the Shannon theorem: beat frequencies measured for detuning pathlengths are bounded by the sampling bandwidth: $\omega_{\rm S} = 2\beta \Delta z/c$. For a wavelength sweep from $\lambda_1$ to $\lambda_2$ during the time $T$, this translates to
\begin{equation}\label{eq_AxialRange}
\Delta z = \frac{N_z}{2} \Lambda 
\end{equation}
where $\Lambda = \lambda_1 \lambda_2/\Delta\lambda = 14.3 \,\mu{\rm m}$ is twice the axial pitch and $N_z = \omega_{\rm S}/\omega_{\rm sweep} = 256$ is the number of interferograms measured during one frequency sweep. The variation of the axial range $\Delta z$ versus wavelength tuning range $\Delta\lambda$ is illustrated experimentally in Fig.~\ref{fig_ScalesOCT}. For tunable lasers with non-Gaussian output spectra~\cite{KleinHuber2017}, the theoretically limiting axial resolution is the round-trip coherence length $\delta z \approx \bar \lambda ^2/ \Delta \lambda = 14.3 \,\mu{\rm m}$, where $\bar \lambda = (\lambda_1 + \lambda_2)/2$ is the central frequency of the sweep. This round-trip coherence length is about the same length as the axial pitch: $\delta z \simeq \Lambda$.

%Relationships in Eq.~\ref{eq_PhaseDelay}, Eq.~\ref{eq_BeatingFrequencyVsPathlength}, and Eq.~\ref{eq_AxialRange} imply that the axial precision of the tomograms relies on sweep range and sampling stability.\\

%
\begin{figure}[]
\centering
\includegraphics[width = 8.0 cm]{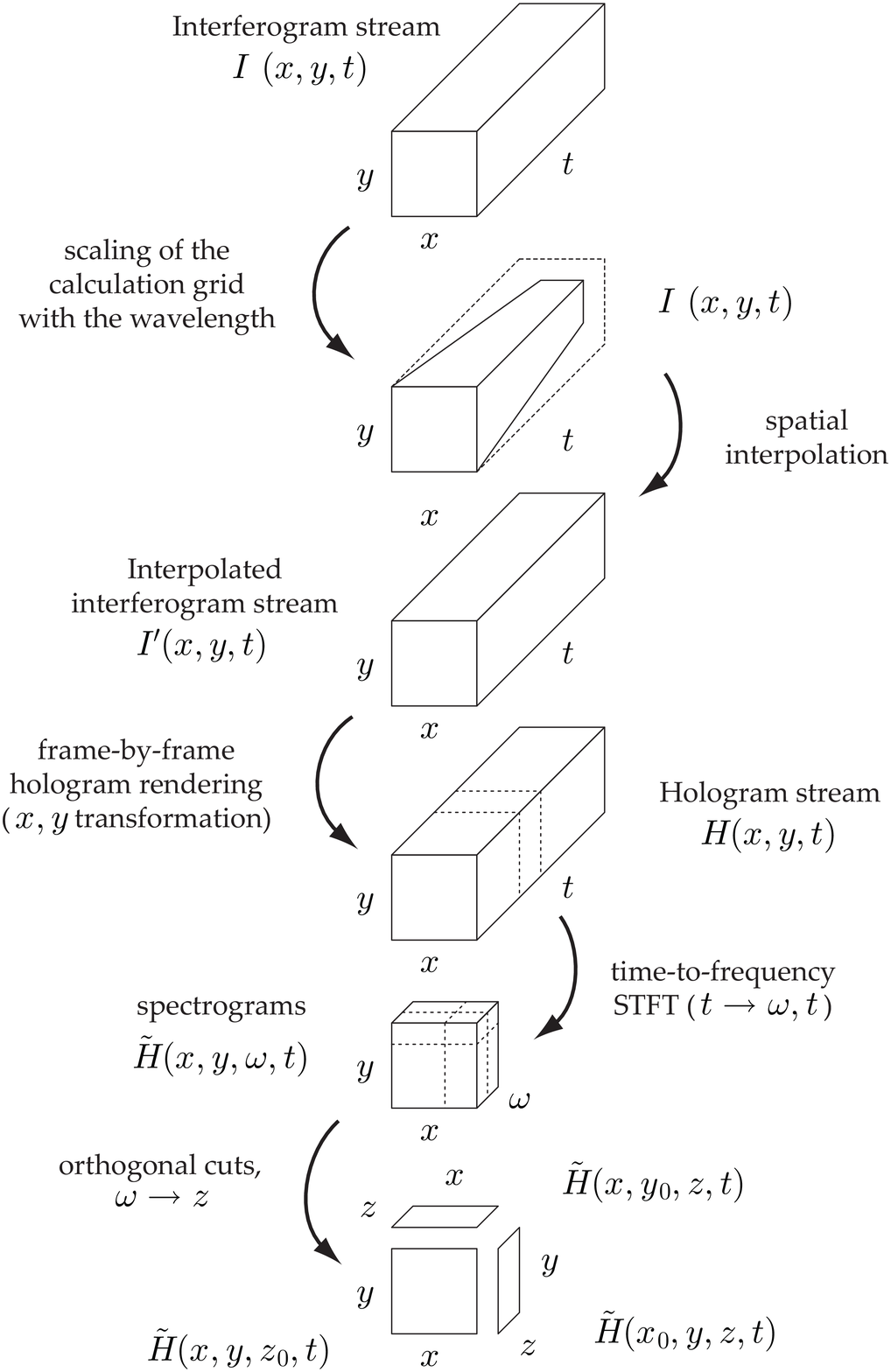}
\caption{Flowchart of the main data processing steps. Each interferogram $I$ from the input data stream is rescaled by linear interpolation to $I'$ to compensate for chromatic distortion. Holograms $H$ are calculated by Fresnel transformation from rescaled interferograms $I'$. Temporal short-time Fourier transforms at each pixel location $(x,y)$ turn holograms $H$ into spectrograms $\tilde{H}$. Orthogonal cuts of the magnitude of each spectrogram $|\tilde{H}|$ are displayed at video rate.
}
\label{fig_STFT_Flowchart}
\end{figure}
\begin{figure}[]
\centering
\includegraphics[width = 8.0 cm]{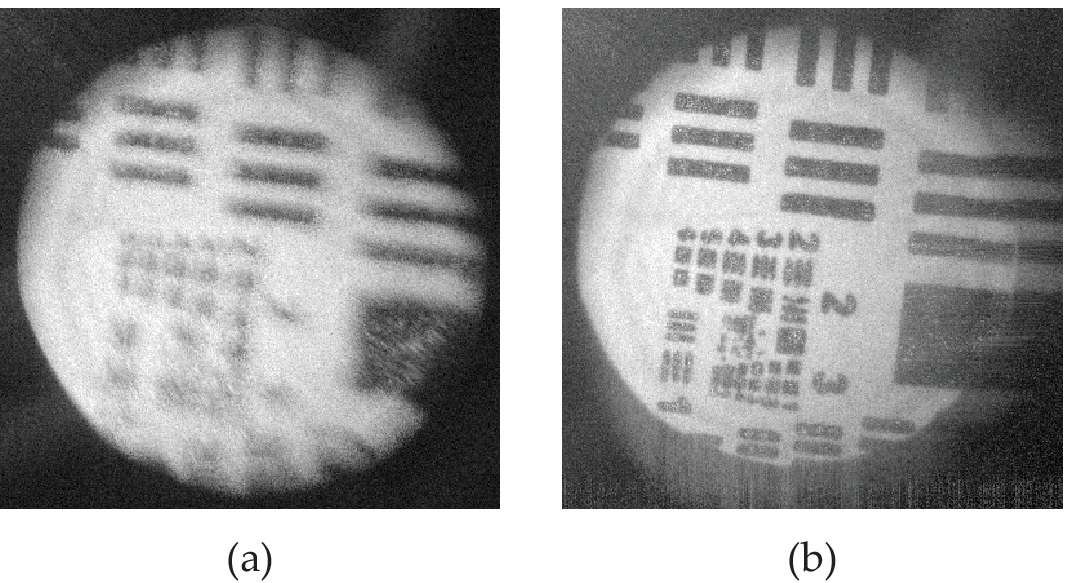}
\caption{illustration of chromatic compensation by interferogram resampling. Hologram rendering without interpolation in \href{https://youtu.be/lwqnLBnSVn4}{Vizualization 1}.}
\label{fig_InterferogramInterpolation}
\end{figure} 

\begin{figure}[]
\centering
\includegraphics[width = 8.0 cm]{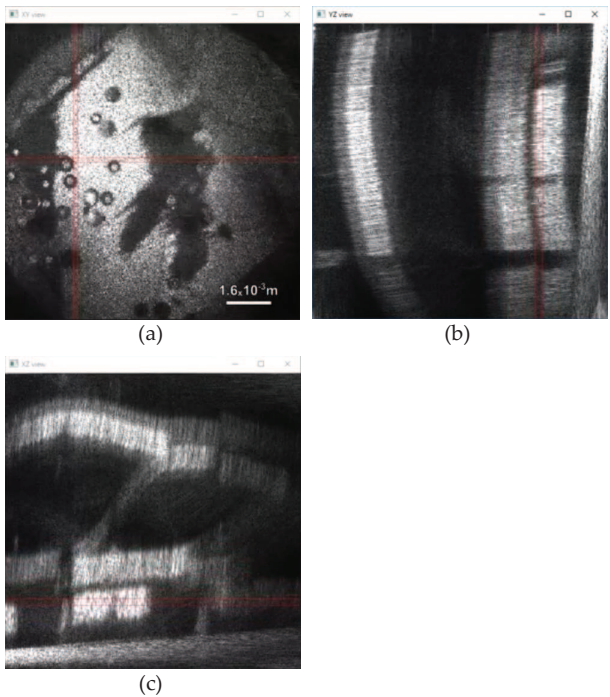}
\caption{Real-time image rendering and visualization in swept-source holographic optical coherence tomography. \href{https://youtu.be/-c6gCsNBppM}{Vizualization 2}. Upper-left window : en-face section. Upper-right and Bottom-left windows : orthogonal B-scans. The measured sample is a static phantom made of 400-800 microns glass beads embeded in several layers of 65 micron-thick tape.}
\label{fig_HolovibesTargetSTFT}
\end{figure}

The sequence of steps reported in Fig.~\ref{fig_STFT_Flowchart} from the input stream of recorded interferograms to the stream of tomographies illustrate the image rendering pipeline. A prerequisite to avoid distortion and loss of resolution of tomograms calculated from Fresnel-transform holograms is to enforce constant values of pixel dimensions in the image plane for all sampling wavelengths of the sweep~\cite{FerraroNicolaFinizio2004, LedlPsotaKavan2017}. In contrast with angular spectrum propagation of the wave field~\cite{YuKim2005, Latychevskaia2015}, the lateral size of a reconstructed pixel of a hologram calculated by Fresnel transformation (Eq.~\ref{eq_FresnelTransform})
\begin{equation}\label{eq_ReconstructedPixelSize}
d' = \frac{\lambda z }{N_x d}
\end{equation}
scales up linearly with the wavelength. This chromatic variation can be counterbalanced by resampling each interferogram with a wavelength-dependent pitch $d_\lambda$. Using a variable sampling pitch $d_\lambda = \lambda d/\lambda_1$ makes the reconstruction pixel $d'$, and hence the lateral field of the reconstructed hologram, wavelength-independent. In practice, each interferogram $I$ was resampled by linear interpolation to $I'$ onto a calculation grid with the same number of points ($N_x \times N_y$) with a pitch $d_\lambda$.  
\begin{equation}\label{eq_InterpolationInterferogram}
I'(x,y,t) = I(x\lambda/\lambda_1,y\lambda/\lambda_1,t)
\end{equation}
Fig.~\ref{fig_STFT_Flowchart} illustrates experimental chromatic stretching compensation by linear interferogram resampling, and Fig.~\ref{fig_InterferogramInterpolation} shows its effect on the lateral resolution of a tomographic hologram of a resolution target for a wavelength sweep from $\lambda_1 = 870$ nm to $\lambda_2 = 820$ nm. Image rendering of complex-valued holograms $H(x,y,t)$ from the stream of rescaled interferograms $I'(x,y,t)$ was performed by discrete Fresnel transformation~\cite{Latychevskaia2015}
\begin{eqnarray}\label{eq_FresnelTransform}
\nonumber H(x,y,t) = \frac{i}{\lambda_1 z}\exp \left( -i k_1 z \right) \iint I'(x',y',t)\\
\times \exp \left[\frac{-i \pi}{\lambda_1 z} \left((x-x')^2 + (y-y')^2\right) \right] {\rm d}x' {\rm d}y'
\end{eqnarray}
where the parameter $z$ corresponds to the sensor-to-object distance for a flat reference wavefront and in the absence of lens in the object path. Eq.~\ref{eq_FresnelTransform} is used for reconstruction parameters $z \geq z_b$, where $z_b = N_x d^2/\lambda_1$. Eq.~\ref{eq_FresnelTransform} is used with parameters $k_1$ and $\lambda_1$ for the reconstruction of all the digital holograms throughout the sweep, regardless of the wavelength to which they correspond, in consequence of wavelength rescaling (Eq~.\ref{eq_InterpolationInterferogram}). Demodulation of the axial depth $z$ of the tomograms consists in forming the beat frequency spectrum of the holograms. For that purpose, temporal short-time Fourier transforms $\tilde{H}(x,y,\omega,t)$ are calculated from the stream of holograms $H(x,y,t)$.
\begin{equation}\label{eq_STFT}
\tilde{H}(x,y,\omega,t) = \int H(x,y,\tau) g(t-\tau) e^{-i \omega \tau} \, {\rm d}\tau  
\end{equation}
where $g(t)$ is a rectangular time gate of $N_z=256$ consecutive images. Then, the envelope of $\tilde H$ is formed, and axial pathlength detunings $z$ are calculated from beat frequencies $\omega$ via Eq.~\ref{eq_BeatingFrequencyVsPathlength}. Three orthogonal cuts $|\tilde{H}(x_0,y,z,t)|^2$, $|\tilde{H}(x,y_0,z,t)|^2$, and $|\tilde{H}(x,y,z_0,t)|^2$ of the magnitude of the rendered volume of a static semi-transparent phantom at arbitrary locations $(x_0,y_0,z_0)$ are displayed in Fig.\ref{fig_HolovibesTargetSTFT}. Swept-source digital holographic OCT was implemented in the software Holovibes (www.holovibes.com). The software has five independent threads dedicated to 1- image acquisition, 2- holographic rendering (spatial demodulation), 3- time-to-frequency analysis (temporal demodulation), 4- image display, and 5- image saving. The acquisition of the raw video stream from the camera is bufferized to avoid any image drop, in order to ensure the consistency of the temporal demodulation (Eq.~\ref{eq_STFT}). The propagation integral in Eq.~\ref{eq_FresnelTransform} is computed by the function \emph{cufft2d()}; the discrete Fourier transform in Eq.~\ref{eq_STFT} used to create local spectrograms is computed by the function \emph{cufft()}. Holovibes is written in C++ and compiled with Microsoft Visual Studio 2017 and NVIDIA CUDA toolkit 9.1, all calculations are performed on 2$\times$32-bit single precision floats per complex value.

\begin{table}[]
%% increase table row spacing, adjust to taste
%\renewcommand{\arraystretch}{1.3}
\label{table_Benchmarks}
\centering
\begin{tabular}{|c|c|c|}
\hline
\, rendering \, & \, input throughput \, & \, output throughput \,  \\
\hline
\hline
\, Fresnel \, & \, 1 GB/s \, & \, 10 Gvoxel/s \, \\
\hline
\, Fresnel + CC \, & \, 512 MB/s \, & \, 10 Gvoxel/s \, \\
\hline
\end{tabular}
\caption{Typical input/output throughput benchmarks with a Titan Xp card. Input data: 16-bit $1024 \times 1024$-pixel digitized interferograms. Fresnel : Fresnel transformation (Eq.~\ref{eq_FresnelTransform}). Time-frequency short-time Fourier transform on 256 frames (Eq.~\ref{eq_STFT}). CC : Chromatic compensation by interferogram resampling (Eq.~\ref{eq_InterpolationInterferogram}).}
\end{table}

In conclusion, we have demonstrated real-time computation and visualization of off-axis Fresnel transform digital holograms from an input stream of 16-bit, 1024-by-1024-pixel interferograms at a rate of 512 frames per second. The resulting stream of holograms was further processed by short-time Fourier transformation to form local spectrograms from 256 frame stacks at the maximum rate of 38 volumes per second. All calculations were performed on single-precision floating-point complex-valued arrays with one NVIDIA Titan Xp graphics card. The resulting software, Holovibes, was used in a swept-source holographic optical coherence tomography experiment for image rendering by Fresnel transformation and chromatic compensation of the pitch variation with wavelength. The reported results demonstrate the scalability of digital holography for high throughput computational volumic imaging in real-time.

%This makes it, to our knowledge, the fastest digital hologram rendering software available in the state of the art

%main throughput bottlenecks. Input is limited by the acquisition system camera, framegrabber, and the computer bus throughput. Output is limited by the GPU bus bandwidth and FLOPs.

%\section{Acknowledgements}

This work was supported by LABEX WIFI (Laboratory of Excellence ANR-10-LABX-24) within the French Program Investments for the Future under Reference ANR-10-IDEX-0001-02 PSL, and European Research Council (ERC Synergy HELMHOLTZ, grant agreement \#610110). The Titan Xp used for this research was donated by the NVIDIA Corporation.

%\bibliographystyle{unsrt}%unsrt%apsrev
%\bibliography{../Biblio/Bibliography}
%\pagebreak

\end{document}